\documentclass[italian,english]{article}
\usepackage[latin9]{inputenc}
\usepackage{textcomp}
\usepackage{amstext}
\usepackage{amssymb}
\usepackage{esint}

\makeatletter

\DeclareRobustCommand{\greektext}{%
  \fontencoding{LGR}\selectfont\def\encodingdefault{LGR}}
\DeclareRobustCommand{\textgreek}[1]{\leavevmode{\greektext #1}}
\DeclareFontEncoding{LGR}{}{}
\DeclareTextSymbol{\~}{LGR}{126}
\newcommand{\lyxmathsym}[1]{\ifmmode\begingroup\def\b@ld{bold}
  \text{\ifx\math@version\b@ld\bfseries\fi#1}\endgroup\else#1\fi}

\newenvironment{lyxlist}[1]
{\begin{list}{}
{\settowidth{\labelwidth}{#1}
 \setlength{\leftmargin}{\labelwidth}
 \addtolength{\leftmargin}{\labelsep}
 }}
{\end{list}}

\makeatother

\usepackage{babel}

\begin{document}

\title{\textbf{Farewell to black hole horizons and singularities? }}
\maketitle
\begin{lyxlist}{00.00.0000}
\item [{\textbf{Authors:}}] \textbf{C. Corda$^{1,a}$, D. Leiter$^{2}$,
H. J. Mosquera Cuesta$^{3,4,5,1,b}$, S. Robertson$^{6,c}$ and R.
E. Schild$^{7,d}$}\end{lyxlist}
\begin{enumerate}
\item International Institute for Theoretical Physics and Advanced Mathematics
Einstein-Galilei, via Santa Gonda 14 - 59100 Prato, Italy; 
\item Deceased;
\item Departamento de Fìsica, Centro de Cièncias Exatas e Tecnològicas (CCET),
Universidade Estadual Vale do Acaraù, Avenida Doutor Guarani 317,
Campus da Cidao, CEP 62.040-730, Sobral, Cearà, Brazil;
\item Instituto de Cosmologia, Relatividade e Astrofìsica (ICRA-BR), Centro
Brasileiro de Pesquisas Fìsicas, Rua Dr. Xavier Sigaud 150, CEP 22290-180,
Urca Rio de Janeiro, RJ, Brazil;
\item International Center for Relativistic Astrophysics Network (ICRANet),
Coordinating Center, Piazza della Repubblica 10, 65122, Pescara, Italy;
\item Department of Physics, Southwestern Oklahoma State University, Weatherford,
OK 73096; 
\item Harvard-Smithsonian Center for Astrophysics, 60 Garden Street, Cambridge,
MA 02138, USA.
\end{enumerate}
\begin{center}
\textit{E-mail:} \emph{$^{a}$cordac.galilei@gmail.com; $^{b}$herman@icra.it;
$^{c}$stanrobertson@itlnet.net; $^{d}$rschild@cfa.harvard.edu}
\par\end{center}
\begin{abstract}
We consider the fundamental issues which dominate the question about
the existence or non-existence of black hole horizons and singularities
from both of the theoretical and observational points of view, and
discuss some of the ways that black hole singularities can be prevented
from forming at a classical level, i.e. without arguments of quantum
gravity. In this way, we argue that black holes could have a different
nature with respect the common belief. In fact, even remaining very
compact astrophysics objects, they could be devoid of horizons and
singularities. 

Our analysis represents a key point within the debate on the path
to unification of theories. As recently some scientists partially
retrieved the old Einstein's opinion that quantum mechanics has to
be subjected to a more general deterministic theory, a way to find
solutions to the problem of black hole horizons and singularities
at a semi-classical level, i.e. without discussions of quantum gravity,
becomes a fundamental framework. \end{abstract}
\begin{quote}
\textbf{This paper is the latest work co-authored by Darryl Jay Leiter,
February 25, 1937 - March 4, 2011}
\end{quote}
The concept of \emph{black hole} (BH) has fascinated scientists long
before the introduction of Einstein's general relativity (see \cite{key-1}
for a historical review). However, an unsolved problem concerning
BHs is the presence of a spacetime singularity in their core. Such
a problem, originally stated in the historical papers concerning BHs
\cite{key-2,key-3,key-4}, was later generalized in the famous paper
by Penrose \cite{key-5}. It is a common opinion that this problem
will be solved when a correct quantum gravity theory is finally constructed
(see \cite{key-6} for recent developments). 

In this review, we discuss fundamental issues which dominate the question
about the existence or non-existence of BH horizons and singularities
from both of the theoretical and observational points of view, and
we analyse some ways to avoid the development of BH singularities
at a semi-classical level which does not require the need for a quantum
gravity theory. On the other hand, at the present time, an absolute
quantum gravity theory, which implies a total unification of various
interactions has not been obtained and, from a historical point of
view, Einstein believed that, in the path to unification of theories,
quantum mechanics had to be subjected to a more general deterministic
theory, which he called generalized theory of gravitation, but he
did not obtain the final equations of such a theory (see for example
the biography of Einstein which has been written by Pais \cite{key-7}).
At present, this point of view is partially retrieved by some theorists,
starting from the Nobel Laureate G. \textquoteright{}t Hooft \cite{key-8}.
However in this context it is a fundamental requirement that solutions
to the problem of BH horizons and singularities be obtained at a semi-classical
level, i.e. without discussions of quantum gravity. An important point
is that it has been recently shown that the true BHs should have $M=0$,
hence the so called BHs could be something else \cite{key-28}.

We begin by noting that in general relativity the Einstein equation
relates the curvature tensor of spacetime on the left hand side to
the energy-momentum tensor in spacetime on the right hand side \cite{key-1}.
Within the context of the Einstein equation the \emph{strong principle
of equivalence} (SPOE) requires that special relativity must hold
locally for all of the laws of physics in all of spacetime as seen
by time-like observers (see Section 2.1 of \cite{key-9}). Hence,
in the context of the SPOE this implies that the frame of reference
of co-moving observers within a gravitationally collapsing object
are required to always be able to be connected to the frame of reference
of stationary observers by special relativistic transformations with
physical velocities which are less than the speed of light. Recently
plausible arguments have been made which support the idea that physically
acceptable solutions to the Einstein equation will only be those which
preserve the SPOE as \emph{a law of nature} in the universe \cite{key-10,key-11,key-12}.
The observable consequence of preserving the SPOE as a \emph{a law
of nature} would be that compact objects which emerge from the process
of gravitation collapse could not have event horizons (EHs) because
their existence would prevent co-moving observers within a gravitationally
collapsing object from being able to be connected to the frame of
reference of stationary observers by special relativistic transformations
with physical velocities which are less than the speed of light. Hence,
as a result of the SPOE, objects having EHs with non-zero mass would
be physically prohibited \cite{key-10,key-11,key-12}. In particular,
the preservation of the SPOE in the Einstein equation would put an
overall constraint on the nature of the non-gravitational physical
elements which go into the energy-momentum tensor on the right hand
side of the Einstein equation. However this constraint would not uniquely
determine the specific form of the non-gravitational dynamics of the
energy-momentum tensor \cite{key-10,key-11,key-12}. For this reason
many different theories can be constructed (e.g. \emph{eternally collapsing
objects} (ECO), \emph{magnetospheric eternally collapsing objects}
(MECO), \emph{nonlinear electrodynamics} (NLED) \emph{objects}, which
preserve the SPOE and hence can generate highly redshifted compact
objects without EHs \cite{key-10,key-11,key-12,key-13}. Since each
of these different SPOE preserving theories have unique observational
predictions associated with the interaction of their non-gravitational
components with the environment of their highly redshifted compact
objects without EHs, the specific one chosen by Nature can only be
determined by astrophysical observations which test these predictions. 

Thus, it is quite important to clearly identify some of the observational
constraints \cite{key-10,key-11,key-12,key-14}. There are two major
classes of BH candidates, both of which seem to be quite consistent
with the possibility that they might be BHs. There are galactic BH
candidates (GBHC) found primarily among the x-ray binary systems and
supermassive active galactic nuclei (AGN). In quiescent states both
are exceedingly faint, consistent with the possibility that they might
not have radiating surfaces \cite{key-10,key-11,key-12,key-14}. For
both classes of quiescent objects, it is plausible, though not proven,
that matter might be accreting to the central object and disappearing
without radiating, as through an EH. In the case of the GBHC, it has
been established that many are more massive that the approximately
3 solar mass allowed for objects supported by internal kinetic or
degeneracy pressure \cite{key-10,key-11,key-12,key-14}. The 3 solar
mass limit in the compact object is determined by relativistic causality,
which requires that the speed of sound must remain less than the speed
of light while kinetic pressures become relativistic. To exceed the
3 solar mass limit, as many GBHC do, they must either be BHs or something
other than cold catalyzed matter. Since gravity can be locally transformed
away it cannot locally determine either the state of matter or its
equation of state. Because of this fact exotic matter objects seem
to be unlikely candidates for GBHC and AGN. On the other hand, a more
likely possibility for GBHC and AGB involves ECO or MECO, which are
hot compact objects supported by internal radiation pressure in a
pair dominated baryon plasma suspended in an Eddington balance which
prevents trapped surfaces leading to EHs from forming. In this context,
in order to satisfy the quiescent luminosity constraints, the surface
radiations from the ECO or MECO objects must be extremely redshifted
\cite{key-10,key-11,key-12,key-14}.

Among galactic nuclei Sagittarius A{*} (Sgr A{*}) has been observed
with sufficient resolution to establish that it is very small, already
less than a few gravitational radii \cite{key-15}. Clusters of stars
packed tightly enough to account for its mass cannot remain stable
for billions of years. Thus, Sgr A{*} seems likely to be a single
object of some kind. The accretion environment of Sgr A{*} should
be capable of supplying about $3*10^{-6}$ solar mass per year in
a Bondi flow into Sgr A{*} \cite{key-10,key-11,key-12,key-14}, but
only about $10^{-10}$ solar mass per year is needed to account for
its bolometric luminosity. Either the Bondi flow is very radiatively
inefficient with most of the mass disappearing through an EH, or there
must be some mechanism that ejects most of the mass. In this regard,
the polarization of radiation from Sgr A{*} constrains the flow that
can get with a few gravitational radii of Sgr A{*} to be less than
about $10^{-9}$ solar mass per year. This very strongly suggests
that an ejection mechanism operates. Finally, if Sgr A{*} would have
a surface that radiated in equilibrium with an accretion rate, the
accretion rate must be less than about $10^{-13}$ solar mass per
year \cite{key-16}.

The MECO model in \cite{key-10,key-11,key-12,key-14} would have a
strong enough magnetic field to eject most of the Bondi flow into
Sgr A{*}. An ECO model with a substantially weaker magnetic field
would presumably have to rely on the same kinds of ejection mechanisms
that have been proposed for BH models. A similar situation applies
to the GBHC. The MECO model provides a strong enough magnetic field
to drive the spectral state switches observed for both neutron stars
and GBHC in x-ray binaries (and also in AGN).  ECO objects with substantially
weaker magnetic fields would have to rely on the same accretion disk
mechanisms that have been proposed for black hole models.

Thorne showed back in 1965 that pure magnetic energy would not collapse
into a BH state \cite{key-17}. It may be that there are stable gravitationally
compact objects that are composed of relatively cool matter and magnetic
fields without being either ECO or MECO as presently understood.  But
such objects would also be subject to the quiescent accretion rate-luminosity
constraints. So it seems that the nature of the surface and the strength
of intrinsic magnetism are crucial issues. \cite{key-10,key-11,key-12,key-14}.

An ECO is a gravitationally compact mass supported against gravity
by internal radiation pressure \cite{key-18}. In its outer layers
of mass, a plasma with some baryonic content is supported by a net
outward flux of momentum via radiation at the local Eddington limit
$L_{Edd}$ given by (note: we work with in natural units in this paper,
i.e. $G=1$, $c=1$ and $\hbar=1,$ while ordinary units will be used
only for numerical results) 

\begin{equation}
L_{Edd,s}=\frac{4\pi M(1+z_{s})}{\kappa}.\label{eq: Eddington}\end{equation}

Here $\kappa$ is the opacity of the plasma, subscript $s$ refers
to the baryon surface layer and $z_{s}$ is the gravitational redshift
at the surface. In general relativity $z$ is given by \cite{key-1}

\begin{equation}
1+z=\frac{1}{\sqrt{1-\frac{2M}{R}}}.\label{eq: redshift}\end{equation}

For a hydrogen plasma, in ordinary units it is $\kappa=0.4\mbox{ }cm^{2}g^{-1}$
and

\begin{equation}
L_{Edd,s}=1.26\times10^{38}m(1+z_{s})\mbox{ }erg\mbox{ }s^{\lyxmathsym{\textminus}1},\label{eq: Eddington ordinary}\end{equation}
 where $m=\frac{M}{M_{\odot}}$ is the mass in solar units. 

Since the temperature at the baryon surface is beyond that of the
pair production threshold and the compactness is large enough to guarantee
a high rate of photon-photon collisions \cite{key-19}, there is a
pair atmosphere further out that remains opaque. The net outward momentum
flux continues onward, but diminished by two effects, time dilation
of the rate of photon flow and gravitational redshift of the photons.
The escaping luminosity at a location, where the redshift is $z,$
is thus reduced by the ratio $\left(\frac{1+z}{1+z_{s}}\right)^{2}$,
and the net outflow of luminosity as radiation transits the pair atmosphere
and beyond is \begin{equation}
L_{net\mbox{ }out}=\frac{4\pi M(1+z)^{2}}{\kappa(1+z_{s})}.\label{eq: Eddington netto}\end{equation}

Finally, as distantly observed where $z\rightarrow0$, the luminosity
is

\begin{equation}
L_{\infty}=\frac{4\pi M}{\kappa(1+z_{s})}.\label{eq: Eddington infinito}\end{equation}

For hydrogen plasma opacity of $0.4\mbox{ }cm^{2}g^{\lyxmathsym{\textminus}1}$
and a typical stellar mass GBHC of $7M_{\odot}$ this equation yields
$L_{\infty}=8.8\times10^{38}/(1+z_{s})\mbox{ }erg\mbox{ }s^{\lyxmathsym{\textminus}1}$
in standard units. But since the quiescent luminosity of a GBHC must
be less than about $10^{31}\mbox{ }erg\mbox{ }s^{\lyxmathsym{\textminus}1},$
we see that it is necessary to have $z_{s}>8.8\times10^{7}$. Even
larger redshifts are needed to satisfy the quiescent luminosity constraints
for AGN. This is extraordinary, to say the least, but perhaps no more
incredible than the $z=\infty$ of a BH. At the low luminosity of
Eq. (3), the gravitational collapse is characterized by an extremely
long radiative lifetime, $\lyxmathsym{\textgreek{t}}$ \cite{key-10,key-18}
given by:

\begin{equation}
\lyxmathsym{\textgreek{t}}=\frac{\kappa(1+z_{s})}{4\pi},\label{eq: tau}\end{equation}

i.e. $\approx4.5\times10^{8}(1+z_{s})\mbox{ }\mbox{ }years$ in ordinary
units.

With the large redshifts that would be necessary for consistency with
quiescent luminosity levels of BHC, it is clear why such a slowly
collapsing object would be called an \textquotedblleft{}\emph{eternally
collapsing object}\textquotedblright{}. 

For $(1+z)>\sqrt{3}$, radiation is impeded by passage through a small
escape cone such that the fraction of radiation that could escape
if isotropically emitted at radius $R$ would only be 

\begin{equation}
f=27\left[\frac{2M}{R(1+z)}\right]^{2}.\label{eq: frazione}\end{equation}

For very large $z$, $R\approx4M$ and $f=\frac{27}{4(1+z)^{2}}.$ 

At the outskirts of the pair atmosphere of an ECO the photosphere
is reached. Here the temperature and density of pairs has dropped
to a level from which photons can depart without further scattering
from positrons or electrons. Nevertheless, the redshift is still large
enough that their escape cone is small and most photons will not travel
far before falling back through the photosphere. If we let the photosphere
temperature, redshift and star's surface be $T_{p}$, $z_{p}$ and
$\sigma$ respectively, the net escaping luminosity is \cite{key-10}

\begin{equation}
L=\frac{27(2M)^{2}4\pi R^{2}\sigma T_{p}^{4}}{R^{2}(1+z_{p})^{2}}=\frac{4\pi M(1+z_{p})^{2}}{\kappa(1+z_{s})}.\label{eq: L escape}\end{equation}

But in the radiation dominated region beyond the photosphere, the
temperature and redshift are related by 

\begin{equation}
T_{\infty}=\frac{T}{1+z},\label{eq: T infinito}\end{equation}

where $T_{\infty}$ is the distantly observed radiation temperature.
Substituting into the previous equation, we obtain 

\begin{equation}
L_{\infty}=(27)4\pi(2M)^{2}\sigma T_{\infty}^{4}=\frac{4\pi M}{\kappa(1+z_{s})},\label{eq: L infinito 2}\end{equation}

and, for hydrogen plasma opacity, by restoring ordinary units it is

\begin{equation}
T_{\infty}=\frac{2.3\times10^{7}}{\left[m(1+z_{s})\right]^{\frac{1}{4}}}\mbox{ }\mbox{ }Kelvins.\label{eq: T infinito 2}\end{equation}

The left equality of Eq. (9) can be written in terms of the distantly
observed spectral distribution, for which the radiant flux density
at distance $R(>6M)$ and frequency $\nu_{\infty}$ would be 

\begin{equation}
F_{\nu_{\infty}}=4\pi^{2}\nu_{\infty}\frac{27(2M)^{2}}{R\left[\exp(\frac{2\pi\nu_{\infty}}{kT_{\infty}})-1\right]}.\label{eq: flux density}\end{equation}

$k$ in Eq. (\ref{eq: flux density}) is the Boltzmann constant.

As previously discussed \cite{key-10,key-12}, if one naively assumes
that the photon support for an ECO originates from purely thermal
processes, one quickly finds that the temperature in the baryon surface
layer would be orders of magnitude higher than the pair production
threshold. The compactness guarantees that photon-photon collisions
would produce numerous electron-positron pairs \cite{key-19}. This
makes the baryon surface a phase transition zone at the base of an
electron-positron pair atmosphere. Drift currents proportional to
$\frac{\vec{g}\times\vec{B}}{B^{2}}$ reactively generate extreme
magnetic fields there. The stability of the Eddington limited MECO
requires that the surface magnetic field, on the Eddington limited
baryon surface of the MECO, will be quantum electrodynamically limited
to have the values about $10^{20}-10^{22}\mbox{ }Gauss$ required
to create a surface density of bound electron-positron pairs in the
baryon plasma \cite{key-20,key-21}. On the other hand the interior
magnetic field strength inside of a stellar mass MECO-GBHC will have
the much smaller equipartition values that which would be expected
from flux compression during stellar collapse. In this context the
Einstein-Maxwell equations at the MECO surface radius ($R\approx4M$)
imply that the ratio of tangential field on the exterior surface to
the tangential field just under the MECO surface is given by \cite{key-10,key-12}

\begin{equation}
\frac{B_{\theta,S_{+}}}{B_{\theta,S_{\lyxmathsym{\textminus}}}}=\frac{(1+z_{s})}{2\ln(1+z_{s})}.\label{eq: B}\end{equation}

In ordinary units the result of Eq. (\ref{eq: B}) is $\mbox{ }\approx\frac{10^{20}\mbox{ }Gauss}{B_{in}\sqrt{7M_{\odot}/M}}$.
We have previously taken $B_{in}=2.5\times10^{13}\mbox{ }Gauss$ as
typical of the interior field that can be produced by flux compression
during stellar gravitational collapse. Using this value the solution
of Eq. (\ref{eq: B}) is $1+z_{s}=5.67\times10^{7}m^{1/2}$. The magnetic
moment of a MECO would be $\lyxmathsym{\textgreek{m}}=1.7\times10^{28}m^{5/2}\mbox{ }Gauss\mbox{ }cm^{3}$
in ordinary units.

This magnetic moment and Eq. (12) give the MECO model a good correspondence
with observations of spectral state switches and the radio luminosities
of jets for both GBHC and AGN \cite{key-10,key-11,key-12}. 

The authors of \cite{key-27} describe the detection of a magnetic
field of $10^{8}$ Gauss that cannot result from frame dragging. It
should be a further support of MECO alternative to BH model.

On the other hand, it has been recently shown that NLED objects can
remove BHs singularities \cite{key-13}.

NLED Lagrangian has been used in various analysis in astrophysics,
like surface of neutron stars \cite{key-22} and pulsars \cite{key-23},
and also on cosmological context to remove the big-bang singularity
\cite{key-24,key-25}. 

The effects arising from a NLED become quite important in super-strongly
magnetized compact objects, such as pulsars and particular neutron
stars \cite{key-22,key-23}. Some examples include the so-called magnetars
and strange quark magnetars. In particular, NLED modifies in a fundamental
basis the concept of gravitational redshift as compared to the well
established method introduced by standard treatments \cite{key-22}.
The analyses proved that, unlike using standard linear electrodynamics,
where the gravitational redshift is independent of any background
magnetic field, when a NLED is incorporated into the photon dynamics,
an effective gravitational redshift appears, which happens to depend
decidedly on the magnetic field pervading the pulsar. An analogous
result has also been obtained for magnetars and strange quark magnetars
\cite{key-23}. The resulting gravitational redshift tends to infinity
as the magnetic field grows larger \cite{key-22,key-23}, as opposed
to the predictions of standard analyses which involve linear electrodynamics.
What it is important is that the gravitational redshift of neutron
stars is connected to the mass\textendash{}radius relation of the
object \cite{key-22,key-23}. Thus, NLED effects turn out to be important
as regard to the mass-radius relation, which is maximum for a BH. 

Following this approach, in \cite{key-13} a particular non singular
exact solution of Einstein field equation has been found adapting
to the BH case the cosmological analysis in \cite{key-25}. In fact,
the conditions concerning the early era of the Universe, when very
high values of curvature, temperature and density were present \cite{key-1,key-13,key-26},
and where matter should be identified with a primordial plasma \cite{key-1,key-13,key-26},
are similar to the conditions concerning BH physics. This is exactly
the motivation because various analyses on BHs can be applied to the
Universe and vice versa \cite{key-1,key-13,key-26}. 

The model works on a homogeneous and isotropic star (a collapsing
{}``\emph{ball of dust}'') supported against self-gravity entirely
by radiation pressure. Let us consider the Heisenberg-Euler NLED Lagrangian
\cite{key-13,key-24,key-25}

\begin{equation}
\mathcal{L}_{m}\equiv-\frac{1}{4}F+c_{1}F^{2}+c_{2}G^{2},\label{eq: 3}\end{equation}

where $G=\frac{1}{2}\eta_{\alpha\beta\mu\nu}F^{\alpha\beta}F^{\mu\nu}$,
$F\equiv F_{\mu\nu}F^{\mu\nu}$ is the electromagnetic scalar and
$c_{1}$ and $c_{2}$ are constants. Through an averaging on electric
and magnetic fields \cite{key-13,key-24,key-25}, the Lagrangian (\ref{eq: 3})
enables a modified radiation-dominated equation of state ($p$ and
$\rho$ are the pressure and the density of the collapsing star) \begin{equation}
p=\frac{1}{3}\rho-\rho_{\gamma},\label{eq: star}\end{equation}

where\emph{ a quintessential density term} $\rho_{\gamma}=\frac{16}{3}c_{1}B^{4}$
is present together with the standard term $\frac{1}{3}\rho$ \cite{key-13,key-24,key-25}.
$B$ is the strength of the magnetic field associated to $F$. The
interior of the star is represented by the well-known Robertson\textendash{}Walker
line-element \cite{key-1,key-13}

\begin{equation}
ds^{2}=-dt^{2}+a(t)[d\chi^{2}+\sin^{2}\chi(d\theta^{2}+\sin^{2}\theta d\varphi^{2})].\label{eq: metrica conformemente piatta}\end{equation}

Using $\sin\chi$ we choose the case of positive curvature, which
is the only one of interest because it corresponds to a gas sphere
whose dynamics begins at rest with a finite radius \cite{key-1,key-13}.
Considering Eq. (\ref{eq: 3}) together with the stress-energy tensor
of a relativistic perfect fluid \cite{key-1,key-13,key-25} 

\begin{equation}
T=\rho u\otimes u-pg,\label{eq: stress energy 2}\end{equation}
 where $u$ is the four-vector velocity of the matter and $g$ is
the metric, the Einstein field equation gives the relation \cite{key-13,key-25}

\begin{equation}
t=\int_{a(0)}^{a(t)}dz(\frac{B_{0}^{2}}{6z^{2}}-\frac{8c_{1}B_{0}^{4}}{6z^{6}}-1)^{-\frac{1}{2}},\label{eq: soluzione}\end{equation}

being $B_{0}=a^{2}B.$ The expression (\ref{eq: soluzione}) is not
singular for values of $c_{1}>0$ in Eq. (\ref{eq: 3}) \cite{key-13,key-25}.
In fact, the presence of the quintessential density term $\rho_{\gamma}$
permits to violate the reasonable energy condition \cite{key-1,key-26}
of the singularity theorems. By using elliptic functions of the first
and second kind, one gets a parabolic trend for the scale factor near
a minimum value $a_{f}$ in the final stages of gravitational collapse
\cite{key-13}.

In concrete terms, by calling $l,m,n$ the solutions of the equation
$8c_{1}B_{0}^{4}-B_{0}^{2}x+3x^{3}=0,$ it is \cite{key-13,key-25}

\begin{equation}
\begin{array}{c}
t=[-(m-l)^{\frac{1}{2}}\beta_{1}(\arcsin\sqrt{\frac{z-l}{m-l}},\sqrt{\frac{l-m}{l-n}})\\
\\+n(m-l)^{-\frac{1}{2}}\beta_{2}(\arcsin\sqrt{\frac{z-l}{m-l}},\sqrt{\frac{l-m}{l-n}})]|_{z=a^{2}(0)}^{z=a^{2}(t)},\end{array}\label{eq: soluzione-1}\end{equation}

where \begin{equation}
\beta_{1}(x,y)\equiv\int_{0}^{\sin x}dz[(1-z^{2})^{-1}(1-y^{2}z^{2})^{-1}]\label{eq: ell1}\end{equation}
 is the elliptic function of the first kind and \begin{equation}
\beta_{2}(x,y)\equiv\int_{0}^{\sin x}dz[((1-z^{2})^{-1})^{-\frac{1}{2}}((1-y^{2}z^{2})^{-1})^{\frac{1}{2}}]\label{eq: ell2}\end{equation}
 is the elliptic function of the second kind.

Then, recalling that the Schwarzschild radial coordinate, in the case
of the BH geometry (\ref{eq: metrica conformemente piatta}), is $r=a\sin\chi_{0}$
\cite{key-1,key-13}, where $\chi_{0}$ is the radius of the surface
in the coordinates (\ref{eq: metrica conformemente piatta}), one
gets a final radius of the star \cite{key-13}

\begin{equation}
r_{f}=a_{f}\sin\chi_{0}>2M\label{eq: non singolare}\end{equation}
 if $B_{0}$ has an high strength, where $M$ is the mass of the collapsed
star and $2M$ the gravitational radius in natural units \cite{key-1,key-13}.
Thus, we find that the mass of the star generates a curved space-time
without EHs.

\subsection*{Conclusion remarks}

In this work we considered the fundamental issues which dominate the
question about the existence or non-existence of BH horizons and singularities
from both of the theoretical and observational points of view, and
discussed some of the ways that BH singularities can be prevented
from forming at a semi-classical level, i.e. without arguments of
quantum gravity. In this way, we argued that BHs could have a different
nature with respect the common belief. Even remaining very compact
astrophysics objects, they could be devoid of horizons and singularities. 

In fact, all of the various models discussed fall into the context
of preserving the SPOE. Which one will be the correct one will ultimately
be determined by their application to observational data on GBHC,
AGN, and the compact object in the center of the galaxy. On this fundamental
issue, we hope in a further improvement of astrophysical observations
in next years.

The discussed framework is a key point within the debate on the path
to unification of theories. As recently some scientists, like the
Nobel Laureate G. \textquoteright{}t Hooft \cite{key-8}, partially
retrieved the old Einstein's opinion \cite{key-7} that quantum mechanics
has to be subjected to a more general deterministic theory, it is
clear that finding solutions to the problem of black hole horizons
and singularities at a semi-classical level, i.e. without discussions
of quantum gravity, becomes a fundamental framework.

\subsection*{Acknowledgements}

The authors thank Abhas Mitra for various useful discussions on the
issues of this paper.

\end{document}